\documentclass[traditabstract]{aa}

\usepackage{epsfig}
\usepackage{graphicx,natbib}
\usepackage{natbib,subfigure} 
\usepackage{amsmath,amssymb}

\def\msun{ M_\odot}
\def\mjup{ M_{\rm J}}
\def\rjup{R_{\rm J}}

\def\beq{\begin{equation}}
\def\eeq{\end{equation}}
\def\simgr{\,\hbox{\hbox{$ > $}\kern -0.8em \lower 1.0ex\hbox{$\sim$}}\,}
\def\simle{\,\hbox{\hbox{$ < $}\kern -0.8em \lower 1.0ex\hbox{$\sim$}}\,}

\def\dd{\mathrm{d}}
\def\ii{\mathrm{i}}

\def\kp{k_{2,\mathrm{p}}}
\def\qp{Q'_{\mathrm{p}}}
\def\Kp{K_{\mathrm{p}}}
\def\dtp{\Delta t_\mathrm{p}}
\def\Rp{R_\mathrm{p}}
\def\Mp{M_\mathrm{p}}
\def\op{\omega_\mathrm{p}}
\def\ep{\varepsilon_\mathrm{p}}
\def\ks{k_{2,\star}}
\def\qs{Q'_{\star}}
\def\Ks{K_{\star}}
\def\dts{\Delta t_\star}
\def\Rs{R_\star}
\def\Ms{M_\star}
\def\os{\omega_\star}
\def\es{\varepsilon_\star}

\titlerunning{Coupled internal/orbital evolution of Hot Jupiters}
\authorrunning{Leconte et al.}

\begin{document}

\title{Is tidal heating sufficient to explain bloated exoplanets?\\ Consistent calculations accounting for finite initial eccentricity}

\author{J\'er\'emy Leconte\inst{1} \and  Gilles Chabrier\inst{1} \and Isabelle Baraffe\inst{1,2} \and  Benjamin Levrard\inst{1}
}

\institute{  \'Ecole Normale Sup\'erieure de Lyon, 46 all\'ee d'Italie, F-69364 Lyon cedex 07, France; \\ 
 Universit\'e Lyon 1, Villeurbanne, F-69622, France; CNRS, UMR 5574, Centre de Recherche Astrophysique de Lyon;\\
 (jeremy.leconte, chabrier, ibaraffe@ens-lyon.fr)
 \and
School of Physics, University of Exeter, Stocker Road, Exeter EX4 4PE, UK}

\offprints{J. Leconte}

\abstract{In this paper, we present the consistent evolution of short-period exoplanets coupling the tidal and gravothermal evolution of the planet. Contrarily to
previous similar studies, our calculations are based on the {\it complete} tidal evolution equations of the Hut (1981) model, valid at any order in eccentricity, obliquity and spin. We demonstrate, both
analytically and numerically, that, except if the system was {\it formed} with a nearly circular orbit ($e\simle 0.2$), solving consistently the complete tidal equations is mandatory to derive correct tidal evolution histories. We show that calculations based on tidal models truncated at second order in eccentricity, as done in all previous studies, lead to quantitatively, and sometimes even qualitatively, erroneous tidal evolutions. As a consequence, tidal energy dissipation rates are severely underestimated in all these calculations and the characteristic
timescales for the various orbital parameters evolutions can be wrong by up to three orders in magnitude. Such discrepancies can by no means be justified by
invoking the uncertainty in the tidal quality factors.

Based on these complete, consistent calculations, we revisit the viability of the tidal heating hypothesis to explain the anomalously large radius of transiting giant planets. We show that, even though tidal dissipation does
provide a substantial contribution to the planet's heat budget and can explain some of the moderately bloated hot-Jupiters, this mechanism can not explain alone the properties of the most inflated objects, including HD 209\,458\,b. Indeed, solving the complete tidal equations shows that enhanced tidal dissipation and thus orbit circularization occur too early during the planet's evolution to provide enough extra energy at the present epoch.
In that case, either a third, so far undetected, low-mass companion must be present to keep exciting the eccentricity of the giant planet, or other mechanisms, such as stellar irradiation
induced surface winds dissipating in the planet's tidal bulges and thus reaching the convective layers, or inefficient flux transport by convection in the planet's interior must be
invoked, together with tidal dissipation, to provide all the pieces of the abnormally large exoplanet puzzle. }


\maketitle

\section{Introduction}
\label{sec:intro}

Gravitational tides have marked out the history of science and astrophysics since the first assessment by Seleucus of Seleucia of the relation between the height of the tides and the position of the moon and the Sun in the second century BC. Modern astrophysics extended the study of gravitational tides in an impressive variety of contexts from the synchronization of the moon and other satellites to the evolution of close binary stars and even the disruption of galaxies.

The recent discoveries of short period extrasolar planetary systems and the determination of the anomalously large radius of some giant close-in exoplanets revived the need for a theory of planetary tides covering a wider variety of orbital configurations than previously encountered for the
case of our own solar system planets. In particular, the orbital evolution of planetary systems such as HD 80\,606, with an orbital eccentricity of 0.9337 \citep{NLM01}, and XO-3, with a stellar obliquity $\es\gtrsim37.3\pm3.7$ deg \citep{WJF09}, cannot be properly treated with tidal models limited to the case of zero or vanishing eccentricity and obliquity such as in the models of e.g. \citet{GS66}, \citet{JGB08} and \citet{FRH08}. 

Following \citet{BLM01} and \citet{GLB03}, attempts have been made to explain the observed large radius of some transiting close-in gas giant exoplanets - the so-called "Hot Jupiters" - by means of tidal heating \citep{JGB08,MFJ09,ISB09}. All these models, however, use tidal models truncated to low (second) order in eccentricity, in spite of initial eccentricities, as determined from the tidal evolution calculations, which can be
as large as $e=0.8$! According to these calculations, a large eccentricity can remain long enough to lead to tidal energy dissipation in the planet's gaseous envelop (assuming a proper dissipation mechanism is at play in the deep convective layers) at a late epoch and then can explain the actual bloated radius of some observed planets. 

In the present paper, we revisit the viability of this tidal heating hypothesis, using an extended version of the \citet{Hut81} tidal evolution model, solving consistently
the {\it complete} tidal equations, to any order in eccentricity and obliquity, and coupling these latter with the gravothermal evolution of the irradiated planet. As will be shown in the paper, properly taking into account the full nature of
the tidal equations severely modifies the planet's tidal and thermal evolution, compared with the aforementioned
truncated calculations, leading to significantly different tidal heat rates and thus planet contraction rates.

After introducing our model in \S\ref{sec:hyp}, we examine in detail in \S\ref{sec:q} the relation between the \textit{constant time lag} ($\Delta t$) in Hut's (and thus our) model and the usual tidal quality factor ($Q$) widely used in the literature. Constraints on $\Delta t$ from the study of the galilean satellites are also derived. In \S\ref{sec:2ndOrder}, we demonstrate, with {\it analytical arguments}, that truncating the tidal equations at $2^{\mathrm{nd}}$ order in eccentricity leads to wrong tidal evolution histories, with sequences drastically differing from the ones obtained when solving the complete equations. In \S\ref{sec:comp}, we compare our full thermal/orbital evolution calculations with similar studies based on a truncated and constant $Q$ tidal model. These numerical comparisons confirm and quantify the conclusions reached
in \S\ref{sec:2ndOrder}, namely that low order eccentricity models substantially underestimate the tidal evolution timescales for initially eccentric systems and thus lead to
incorrect tidal energy contributions to the planet's energy balance. For instance, we show that tidal heating can not explain the radius of HD~209\,458\,b, for the present values of their orbital parameters,
 contrarily to what has been claimed in previous calculations based on truncated eccentricity models \citep{ISB09}. Finally, in \S\ref{sec:global}, we apply our model to the case of some of the discovered bloated planets. We show that, although tidal heating can explain the presently observed
 radius of some {\it moderately bloated} hot Jupiters, as indeed suggested in some previous studies, tidal heating alone cannot explain {\it all} the anomalously
 large radii. Indeed, in these cases, eccentricity damping occurs too early in the system's tidal evolution (assuming a genuine two-body planetary system) to lead to the present state of the planet's contraction.


\section{Model Description}
\label{sec:hyp}

\subsection{Internal evolution}
\label{sec:intevolution}

The main physics inputs (equations of state, internal composition, irradiated atmosphere models, boundary conditions) used in the present calculations have been described in details in previous papers devoted to the evolution of extrasolar giant planets \citep{BCB03,BCB08,LBC09} and
are only briefly outlined below. 
The evolution of the planet is based on a consistent treatment between the outer non-grey irradiated atmospheric structure and  the inner structure. The interior is composed primarily of a gaseous H/He envelope whose thermodynamic properties are described by the Saumon-Chabrier-VanHorn equation of state (EOS,
\citealt{SCV95}) with a solar or non-solar enrichment in heavy elements described by the appropriate EOS's \citep{BCB08}. In the present calculations, our fiducial model consists of  a planet with a central core made up of water, with the ANEOS EOS \citep{TL72}. A detailed analysis
of the effects of different EOS's, core compositions and heavy material repartitions within the planet can be found in \citet{BCB08}, as well as a
comparison with models from other groups, in particular the ones by \citet{FMB07}. 

Transiting planets are by definition very close to their host star ($a< 0.1$AU).
In that case, the stellar irradiation strongly affects the planet atmospheric structure to deep levels \citep{BHA01}
and thus the planet's evolution (\citealt{GBH96}, \citealt{BCB03}, \citealt{BSH03}, \citealt{CBB04}). 
We use a grid of irradiated atmosphere models based on the calculations of
\cite{BHA01}, computed for different levels of stellar irradiation relevant
to the present study. For planets with a finite orbital eccentricity, the mean stellar flux received is given by:
\begin{equation}
\label{finc}
<F_{\mathrm{inc}}> =f\, R_\star^2\  \sigma  T_{\mathrm{eff},\star}^4<\frac{1}{r^2} >= f\,\left( \frac{R_\star}{a } \right)^2 \frac{\sigma T_{\mathrm{eff},\star}^4}{\sqrt{1-e^2}}\,,
\end{equation}
where $\Rs$ and $T_{\mathrm{eff},\star}$ are the stellar radius and effective temperature, $\sigma$ the Stefan-Boltzmann constant, $a$ and $e$ the orbital semi-major axis and eccentricity and $f$ is a geometrical factor (substituting $f$ by 1 in Eq.\,\ref{finc} gives the mean flux received at the substellar point). The atmospheric models were computed with $f=1/2$ as described by \citet{BHA01}.

\subsection{Tidal Model}\label{sec:tidalmodel}

We consider the gravitational tides raised by both the host star and the planet on each other and follow the traditional ``viscous'' approach of the
{\it equilibrium tide} theory \citep{Dar08}. The secular evolution of the semi-major axis $a$ can be calculated exactly 
(e.g. \citealt{Hut81}; \citealt{NL97}; \citealt{LCC07}; \citealt{CL10}; see appendix \ref{appendix} for the derivation of these equations for any value of the eccentricity and obliquity)
\begin{align}
\frac{1}{a}\frac{\dd a}{\dd t}=\frac{4\,a}{GM_{\star}\Mp}
&\Big\{&\Kp&\left[N(e)\,x_\mathrm{p}\,
\frac{\op}{n}-N_a(e)\right]  \nonumber\\
&+& \Ks&\left[N(e)\,x_\star\,
\frac{\os}{n}-N_a(e)\right]\Big\}\ , \quad
\label{evol_a}
\end{align}
with
\begin{equation}
\label{n_e}
N(e) = \frac{1+\frac{15}{2}e^2+\frac{45}{8}e^4+\frac{5}{16}e^6}{(1-e^2)^{6}}
\end{equation}
and
\begin{equation}
\label{na_e}
N_a(e)=\frac{1+\frac{31}{2}e^2+\frac{255}{8}e^4+\frac{185}{16}e^6+\frac{25}{64}e^8}{(1-e^2)^{15/2}},
\end{equation}
where $G$ is the gravitational constant, $\op$ is the planet's rotation rate, $\Mp$ and $\Rp$ its mass and radius, $\ep$
its obliquity (the angle between the equatorial and orbital planes), with $x_\mathrm{p}=\cos
\ep$, and $n$ the orbital mean motion. 
\begin{equation}
\Kp = \frac{3}{2} \kp \dtp  \left( \frac{G
\Mp^2}{\Rp} \right) \left(\frac{\Ms}{\Mp} \right)^2 \left( \frac{\Rp}{a}
\right)^6 n^2 \ ,
\end{equation}
where $\kp$ and $\dtp$ are the potential Love number of degree 2 and the constant time lag for the planet. The stellar parameters correspond to the same definitions, by simply switching the p and $\star$ indices.
Similarly, the secular evolution of the eccentricity is given by
\begin{align}
\frac{1}{e}\frac{\dd e}{\dd t}=11\frac{a}{GM_{\star}\Mp}
&\Big\{&\Kp&\left[\Omega_e(e)\,x_\mathrm{p}\,
\frac{\op}{n}-\frac{18}{11}N_e(e)\right]  \nonumber\\
&+& \Ks&\left[\Omega_e(e)\,x_\star\,
\frac{\os}{n}-\frac{18}{11}N_e(e)\right]\Big\}\ , \quad
\label{evol_e}
\end{align}
with
\begin{equation}
\label{omega_e_e}
\Omega_e(e) =\frac{1+\frac{3}{2}e^2+\frac{1}{8}e^4}{(1-e^2)^{5}}
\end{equation}
and
\begin{equation}
\label{ne_e}
N_e(e)=\frac{1+\frac{15}{4}e^2+\frac{15}{8}e^4+\frac{5}{64}e^6}{(1-e^2)^{13/2}}.
\end{equation}
The terms proportional to $\Kp$ (resp. $\Ks$) are due to the tides raised by the star (planet) on the planet (star).
Finally, the evolution of the rotational state of each object $i$ ($i$ being $p$ or $\star$) is given by:
\begin{equation}
 \frac{\dd C_\ii\omega_\ii}{\dd t}=-\frac{K_\ii}{n}
 \left[\left(1+x_\ii^2\right)\Omega(e)\frac{\omega_\ii}{n}-2x_\ii\,N(e)
 \right] \ ,
\label{rot_tidal}
\end{equation}
while the evolution of the obliquity obeys the equation
\begin{equation}
\frac{\dd \varepsilon_\ii}{\dd t}=\sin \varepsilon_\ii \frac{K_\ii}{C_\ii \omega_\ii\,n}
  \left[(x_\ii-\eta_\ii)\,\Omega(e) \frac{\omega_\ii}{n}- 2\,N(e) \right] \ , \label{rot_tidal2}
\end{equation}
where $C_\ii$ is the principal moment of inertia of the deformable body under consideration, $\eta_\ii$ is the ratio of rotational over orbital angular momentum $$\eta_\ii=\frac{\Mp+\Ms}{\Mp\Ms} \frac{C_\ii \omega_\ii}{a^2n \sqrt{1-e^2}},$$ and 
\begin{equation}
\label{omega_e}
\Omega(e) = \frac{1+3e^2+\frac{3}{8}e^4}{(1-e^2)^{9/2}}.
\end{equation}
Up to this point, no assumption has been made on the objects themselves. As a result, Eqs.\,(\ref{evol_a})-(\ref{rot_tidal2}) are fully symmetric in p and $\star$ indices and can be used directly to model binary stars or a planet-satellite system.
For typical HD 209\,458\,b-like parameters and
$\Delta t$-values comparable to that inferred for Jupiter (see \S\ref{sec:q}), the planetary spin evolves to a coplanar ($\ep=0$) state with the equilibrium rotation rate value (setting $ \dd \op / \dd t = 0$)
\begin{equation}
\omega_{\mathrm{eq}}=\frac{N(e)}{\Omega(e)}\frac{2x}{1+x^2}\,n= \frac{N(e)}{\Omega(e)}\,n\,,
\label{rot_eq}
\end{equation}
within a time scale $\tau=C_\mathrm{p}\,n^2/\Kp \sim 10^5$~yr. Since the age of the known transiting systems ranges from a few Myr to Gyr, we can safely assume that the planet is now in this state of pseudo synchronization.
The evolution of the orbital parameters of the planet-star
systems is thus fully defined by the system of differential equations, to be solved consistently, given by
Eqs.\,(\ref{evol_a}) and (\ref{evol_e}) for $a$ and $e$, 
Eqs.\,(\ref{rot_tidal}) and (\ref{rot_tidal2}) for the stellar rotation and $\ep=0$, $\op=\omega_{\mathrm{eq}}$.
The rate of tidal dissipation within the {\it planet} in this state of pseudo synchronization reads \citep{Hut81, LCC07}
\begin{equation}
\dot{E}_{\mathrm{tides}}=
2\Kp\left[N_a(e)-\frac{N^2(e)}{\Omega(e)} \right] \ ;\ \ (\op=\omega_{\mathrm{eq}})\ .
\label{tidal_energy}
\end{equation}
The dissipated heat is deposited over the whole planet's interior. 

One can see from Eq.\,(\ref{appendtidalnrj2}) in Appendix \ref{appendix} that Eq.\,(\ref{tidal_energy}) is a special case of energy dissipation for a body in pseudo synchronous rotation as expected for fluid objects ($\ep=0$, $\op=\omega_{\mathrm{eq}}$). For a rocky planet, the external gravitational potential created by its permanent quadrupole moment can cause its locking into synchronous rotation ($\op=n$) and the dissipation rate reads in that case
\begin{align}
\dot{E}_{\mathrm{tides}}&=&\,2\,\Kp\ \times\ &\nonumber\\  
&\times&\Bigg[N_a(e)-2&N(e)\,x_\mathrm{p}+\left(\frac{1+x_\mathrm{p}^2}{2}\right)\Omega(e)\Bigg] ;\ (\op=n)\ .
\end{align}
This equation fully agrees with Eq.\,(30) of \citet{Wis08} who calculated it for a homogeneous, incompressible with a radial displacement Love number $h_2 =5k_2/3$. Note that our derivation does not require such an hypothesis and all the uncertainties in the radial distribution of material and its physical properties (e.g., density, compressibility, elasticity) are lumped into the $k_2$ parameter \citep{Lev08}.


\section{Relationship between the time lag $\Delta t$ and the quality factor ($Q$)}
\label{sec:q}

The aforedescribed tidal model, which leads to {\it exact} tidal evolution equations in the viscous approximation, implies a {\it constant time lag} $\Delta t$. Neither the tidal quality factor ($Q$), or its counterpart, the phase lag ($\epsilon$) \citep{Gol63,GS66} enter the dynamical evolution equations. Instead, the model is characterized by the time lag between the maximum of the tidal potential and the tidal bulge in each body, $\Delta t_\ii$, considered to be constant during the evolution. As shown e.g. by \citet{Dar08} (see also \citealt{Gre09}), this model is equivalent to considering a body whose rheology entails $Q^{-1}(\sigma)\approx \epsilon(\sigma)\propto \sigma$, where $\sigma/2\pi$ is the frequency of the tidal forcing. The actual rheology of giant gaseous planets being poorly constrained, this arbitrary choice based on the visco-elastic model has the advantages of (i) not introducing any discontinuity for vanishing tidal frequencies, as is the case for synchronous rotation, and (ii) to allow for a complete calculation of the tidal effect {\it without any assumption on the eccentricity} for an ideal viscoelastic body.

Indeed, as shown by \citet{Gre09}, the frequency dependence of the phase lag of a perfect viscoelastic oscillator is given by 
\begin{equation}
\tan (\epsilon)=\frac{\sigma}{\tau (\omega_0^2-\sigma^2)}
\end{equation}
where $\tau$ is a viscous damping timescale and $\omega_0$ the natural frequency of the oscillator. In an incompressible gaseous body, the restoring force acting against the tidal deformation is the self gravity of the body. Thus $\omega_0$ can be estimated through the free-fall time as $2\pi/\omega_0\approx\frac{1}{4}\sqrt{\frac{3\pi}{2\,G\bar{\rho}}}\approx 30$ minutes for Jupiter mean density. For tidal periods of several days $\omega_0\gg\sigma$, and for weakly viscous fluid, the phase lag reads
\begin{equation}
\epsilon(\sigma)\approx\frac{\sigma}{\tau\, \omega_0^2}\equiv\,\sigma\, \Delta t
\end{equation} 
which is the frequency dependence corresponding to the constant time lag model.

On the contrary, constant-$Q$ models described by \citet{GS66}, \citet{JGB08}, \citet{FRH08} were derived using perturbative developments of Kepler equations of motion both in eccentricity and inclination. Such Fourier decomposition is in fact necessary in an "lag and add" approach with a given frequency dependence of the phase lag ($\epsilon(\sigma)$). Indeed, in this approach, one must first separate the forcing potential in terms with a defined frequency before lagging them with the chosen $\epsilon(\sigma)$ (See \citealt{FRH08,Gre09}). As a result they can only be used in the $e\ll1$ and $\varepsilon_\ii\ll1$ limit.

The time lag $\Delta t$ can be linked to the reduced quality factor $Q'\equiv 3Q/2k_2$, chosen so that $Q'=Q$ for an homogeneous sphere ($k_2=3/2$). Indeed, one must remember that the {\it phase lag}, $\epsilon(\sigma)$, induced by the
tidal dissipative effects, is twice the geometrical lag angle, $\delta(\sigma)$, between the maximum of the deforming potential and the tidal bulge: $\epsilon(\sigma)=2\delta(\sigma)=\sigma\Delta t $. Moreover, for an incompressible body, a reasonable assumption for giant planets, the tidal dissipation function is given by \citep{Gol63,EW09}:
\begin{equation}
Q^{-1}(\sigma)=-\frac{\Delta_{\mathrm{cycle}} E(\sigma)}{2\pi E_{\mathrm{peak}}(\sigma)}=\frac{\tan\epsilon(\sigma)}{1-(\frac{\pi}{2}-\epsilon(\sigma))\tan\epsilon(\sigma)},
\label{Qfunc}
\end{equation}
where $\Delta_{\mathrm{cycle}}E(\sigma)$ is the energy dissipated by the body at the frequency $\sigma$ during one tidal period and $E_{\mathrm{peak}}(\sigma)$ the maximum energy stored in the perturbation. In the cases of interest in the present study, where $Q\gg1$, and for non-synchronized circular orbits, semi-diurnal tides dominate and one can equal the average tidal quality factor to the one given by Eq.\,(\ref{Qfunc}) for the relevant frequency
\begin{equation}\label{QforJ}
Q'^{-1}\approx \frac{4}{3} k_2\Delta t |\omega-n|.
\end{equation}
This formula can be used to estimate the quality factor in the case of the jovian planets as long as semi-diurnal tides dominate. 
As the planet tends toward synchronization, the dissipative effects of the semi diurnal tides ($\sigma=2|\op-~n|$) vanish with their frequency. Then, the most dissipative tides are the eccentric annual tides ($\sigma=n$) and
\begin{equation}
\label{QforExo}
Q_{\mathrm{p}}'^{-1}\approx \frac{2}{3}k_2\Delta t\, n.
\end{equation}
Apart from these two limit cases, no tidal frequency dominates, and the dissipation is the response of the body to the rich spectrum of exciting tidal frequencies. Thus no simple relation exists between $Q'$ and $\Delta t$ in the general case.

Although it is tempting to use Eq.\,(\ref{QforExo}) to rewrite the tidal equation and to keep $Q'$ constant instead of $\Delta t$ as done by, for example, \citet{ML02}, \citet{DLM04} and \citet{BO09}, one must keep in mind that this procedure is not equivalent either to the constant phase lag (i.e. constant $Q$) or  time lag model. Indeed the frequency dependence of the phase lag is given by $\epsilon(\sigma)=\sigma/(nQ)$ and is still proportional to the tidal frequency over an orbit as in the constant time lag model, but with a slope that is changing during the evolution.

In \S\ref{sec:2ndOrder} and \S\ref{sec:comp}, we compare the constant time lag model with the constant $Q'$ model used by various authors. In order to allow a direct and immediate comparison with these studies, we will choose the values of the couple $(\dtp,\dts)$ from the relations
 $(k_2\dtp=\frac{3}{2 n_{\mathrm{obs}} \qp} ,k_2\dts=\frac{3}{2 n_{\mathrm{obs}} \qs} )$, where $(\qp,\qs)$ are the \textit{constant} normalized
 quality factors used by \citet{MFJ09}. This ensures that the effective tidal dissipation function is the same in both calculations for a given planet with its measured orbital parameters.

In order to use the constant \textit{time lag} model, we must consider many values for $\Delta t$. To constrain this parameter, we follow the analysis of \citet{GS66} and use the Io-Jupiter system to infer an upper limit for $\kp\times\dtp$ in giant extrasolar planets. Since Jupiter is rapidly rotating, 
with $\omega_{\mathrm{J}}>n$ (hereafter, J indices refer to the value for Jupiter), where $n$ is the orbital mean motion of Jupiter's satellites, tidal transfer of angular momentum drives the satellites of Jupiter
{\it outwards}, into expanding orbits. Therefore, the presence of Io in a close orbit provides an upper limit for the time lag in Jupiter. Indeed, if $\dtp$ was too large, the backward evolution of the satellites orbits would imply their disappearance within less time than the age of the Solar system, i.e. of Jupiter.
%
%
For coplanar and circular orbits, a dimensionless version of Eq.\,(\ref{evol_a}) reads:
\begin{equation}
\label{adotozero}
\dot{\tilde{a}}=- \frac{1}{\tilde{a}^7}\left[1 -\frac{\os}{n}\right],
\end{equation}
where $\tilde{a}=a/a_0$, 0 indices refer to initial values and time is counted in units of $$\tau = \frac{1}{6} \frac{\Mp^2}{M_\mathrm{S}(M_\mathrm{S}+\Mp)}(\frac{a_0}{\Rp})^8 \frac{\Rp^3}{G\Mp \kp\dtp},$$
where $M_\mathrm{S}$ is the satellite mass. $\tau$ is the typical timescale of tidal evolution of the semi-major axis.
Injecting angular momentum conservation $$J_\mathrm{tot}=\frac{\Mp M_\mathrm{S}}{\Mp+M_\mathrm{S}} a^2n +\sum_\ii C_\ii \omega_\ii$$ into Eq.\,(\ref{adotozero}), and integrating over time yields
\begin{equation}
\label{exactsole0}
\int_1^{\frac{a(t)}{a_0}} \frac{-\tilde{a}^7 \dd \tilde{a}}{1 -\tilde{a}^{3/2}[\frac{\omega_{\mathrm{p},0}}{n_0}-\beta(\sqrt{\tilde{a}}-1)]}=\frac{t}{\tau},
\end{equation}
where 
$\beta =(M_\mathrm{S}\Mp a_0^2)/((\Mp+M_\mathrm{S})C_\mathrm{p})$. Note that this integral cannot be performed down to $a=0$ because the satellite first crosses the corotation radius where the integrand tends to infinity, which is an unstable equilibrium state for the system \citep{Hut80}. Furthermore, this result is not limited to the case of a planet-satellite system and can be directly used to compute the inspiral time of a close-in exoplanet (setting p$\,\rightarrow\star$ and S$\,\rightarrow\,$p).

For the Io\,-\,Jupiter system, taking $t=-4.5\,\times10^9$\,yr and $a(t)$ equal to the Roche limit in Eq.\,(\ref{exactsole0}) yields $k_\mathrm{J}\Delta t_\mathrm{J}\lesssim5\times 10^{-3}$s.
Therefore, for the actual Io\,-\,Jupiter system, Eq.\,(\ref{QforJ}) implies $Q'_\mathrm{J}\gtrsim 1\times10^6$, slightly smaller than the value derived by \citet{GS66}. As discussed by these authors, our upper limit on $\dtp$ must be multiplied by a factor 5 to 7.5, as Io might have been trapped in a low order commensurability with Europa and Ganymede during part of its evolution, slowing down the expansion of its orbit. This roughly yields $$k_\mathrm{J}\Delta t_\mathrm{J}\lesssim2-3\times 10^{-2}\,\mathrm{s}.$$

For sake of easy comparison, we will refer to the quantity $Q'_0$, which is the reduced quality factor computed for a reference period of 1 day:
$$Q'_0=\frac{3}{2}\frac{Q(2\pi/1\,\mathrm{day})}{k_2}=\frac{3}{2}\frac{1\,\mathrm{day}}{2\pi\, k_2 \Delta t}.$$ The above calculated constraint reads $Q'_{0,\mathrm{p}}\gtrsim 1\times10^6$. In the present calculations, we will examine two cases for the planet under consideration, namely $Q'_{0,\mathrm{p}}=10^6$ and $Q'_{0,\mathrm{p}}=10^7$ ($\kp\dtp\sim2\times10^{-2}-2\times10^{-3}$), while taking $Q'_{0,\star}$ in the range $10^5-10^6$ ($\ks\dts\sim2\times10^{-1}-2\times10^{-2}$), a typical value for solar-type stars \citep{OL07}.

It is important to stress that, if $\Delta t$, or its counterpart $Q$, is poorly known for both planets and stars, its variability from one object or configuration to another is even more uncertain. For instance, tidal dissipation in planets probably differs significantly from the one in brown dwarfs
because of the presence of a dense core able to excite inertial waves in the convective envelope \citep{GL09}. Given the highly
non-linear behavior of tidal dissipation mechanisms, the effective tidal dissipation function varies not only with the structure of the object or with
the tidal frequency but also with the amplitude of the tidal potential. For example, $\qs$ values inferred from the circularization of close FGK binary stars \citep{MM05}, may be lower than the actual $\qs$ encountered in star-planet systems \citep{OL07}. Consequently, the range of values considered here for both $\qs$ and $\qp$ should be seen as mean values and be re-evaluated when considering specific and/or atypical systems (XO-3, HAT-P-2 or CoRoT-Exo-3 for example).


\section{Effect of the truncation of the tidal equations to $\mathbf{2^{\mathrm{nd}}}$ order in $\mathbf{e}$: Analytical analysis.
}\label{sec:2ndOrder}

Following the initial studies of \citet{JGB08}, all the studies exploring the effect of tidal heating on the internal evolution of "hot jupiters" (\citealt{MFJ09}, \citealt{ISB09}) have been using a tidal model assuming a constant $Q$ value during the evolution. Moreover, in all these calculations, the tidal evolution
equations are truncated at the second order in eccentricity (hereafter referred to as the "$e^2$ model"), even when considering tidal evolution
sequences with non-negligible values of $e$ at earlier stages of evolution. Although such a $e^2$-truncated model is justified for planets and satellites in the solar system \citep{Kau63,GS66}, it becomes invalid, and thus yields incorrect results for $a(t)$, $e(t)$ and $\dot{E}_{\mathrm{tides}}$ for finite eccentricity values. 
The main argument claimed for using such a simple tidal model is the large uncertainty on the tidal dissipation processes in astrophysical objects. In particular, as detailed by \citet{Gre09}, the linearity of the response to the tidal forcing based on the viscoelastic model may not hold in a real object for the large spectrum of exciting frequencies encountered when computing high order terms in the eccentricity. Although the large uncertainty in the dissipative processes certainly precludes an exact determination of the tidal evolution, it can by no means justify calculations which are neglecting dominant terms at finite $e$.

Indeed, from a dimensional point of view and prior to any particular tidal model, the strong impact of high order terms in the eccentricity is simply due to the fact that the tidal torque ($\mathbf{N}$) is proportional to $(\omega-\dot{\theta})/r^{6}$ ($\theta$ being the true anomaly) and that over a keplerian orbit, the average work done by the torque is of the form $$<\mathbf{N}\cdot\mathbf{\dot{\theta}}>\,\propto\,<\frac{\dot{\theta}^2}{r^6}>=\frac{n^2}{a^6}\cdot\frac{1+14e^2+\frac{105}{4}e^4+\frac{35}{4}e^6+\frac{35}{128}e^8}{(1-e^2)^{15/2}},$$ which is a rapidly increasing function of $e$ (see Appendix \ref{appendix} for the details of the calculation). This means that, although the mean distance between the planet and the star increases with $e$, the distance at the periapsis strongly {\it decreases}, and most of the work due to the tidal forces occurs at this point of the orbit. One can see that for $e>0.32$ the high order terms dominate the constant and $e^2$ terms. This is a physical evidence that shows that for moderate to large eccentricity, most of the tidal effects are contained in the high order terms that can therefore not be neglected independently of any tidal model.


In this section we quantify more comprehensively this statement. We will demonstrate analytically that:
\begin{itemize}
\item in the context of the Hut model, a truncation of the tidal equations at the order $e^2$ can lead not only to quantitatively wrong but to
{\it qualititatively} wrong tidal evolution histories, with sequences drastically differing from the ones obtained with the complete solution.
\item the rate of tidal dissipation can be severely underestimated by the quasi circular approximation ($e\ll1$).
\end{itemize}
Furthermore, $Q$-constant models consider only low order terms in obliquity ($\varepsilon_\ii$), and thus cannot address the problem of obliquity tides and energy dissipation produced by this mechanism. A detailed discussion on this subject is presented in \citet{LCC07} and \citet{BO09} and will not be reproduced here.

\subsection{Expanding vs shrinking orbits}

\begin{figure}[htbp]
\begin{center}
 \subfigure[$\omega/n$ vs $e$ phase space: complete equations]{ \includegraphics[width=8.5cm]{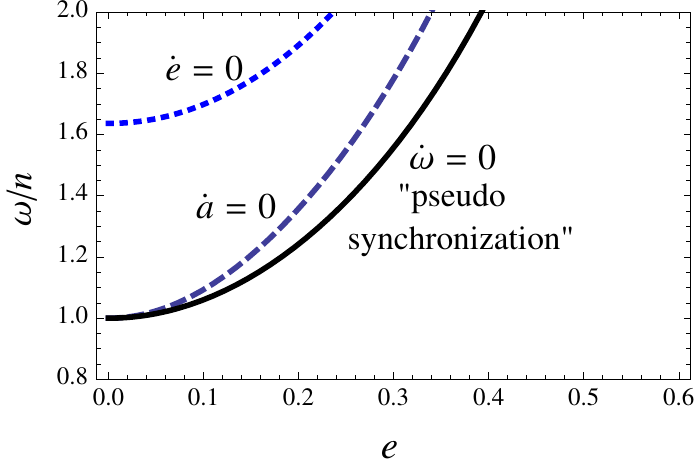} }
 \subfigure[$\omega/n$ vs $e$ phase space: truncated equations]{ \includegraphics[width=8.5cm]{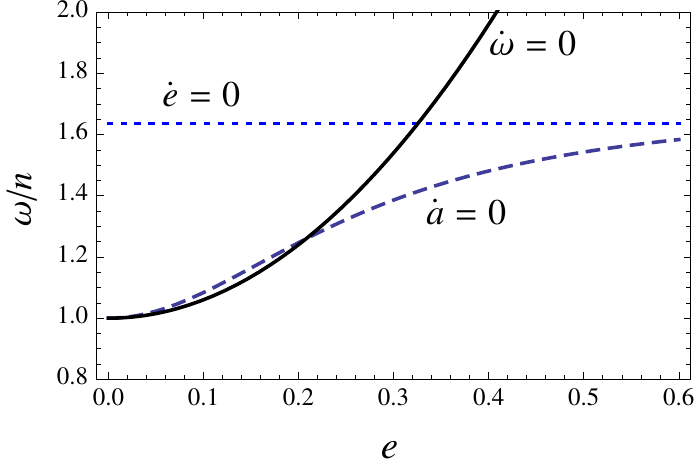}} \\
\end{center}
\caption{Pseudo synchronization curve (\textit{solid}), the $\dot{a}=0$ curve (\textit{dashed}) and the $\dot{e}=0$ curve (\textit{dotted}) for the complete model (top pannel) and for the truncated one (bottom pannel) in the $\omega/n$ vs $e$ phase space. A pseudo synchronized planet always follows the $\dot{\omega}=0$ curve (black solid curve). It always lies in the $\dot{a}<0$ and $\dot{e}<0$ part of the diagram because when solving the complete Hut tidal equations, these curves do not intersect (top pannel). In contrast, as demonstrated in \S\ref{sec:2ndOrder}, in the $2^{\mathrm{nd}}$-order truncated model, the pseudo synchronization curve intersects the $\dot{a}=0$ (at $e\sim0.208$) and $\dot{e}=0$ lines ($e\sim0.326$).
}
\label{fig:phasespace}
\end{figure}

On one hand, considering Eq.\,(\ref{evol_a}) (with $\varepsilon=0$ for simplification) we can see that, for $\omega_\ii/n\leqslant N_a(e)/N(e)$, the tides raised on the body $i$ lead to a decrease of the semi-major axis, transferring the angular momentum from the orbit to the body's internal rotation. It is easy to show that for a synchronous planet this condition is always fulfilled, since $\frac{\omega_\mathrm{eq}}{n}=\frac{N(e)}{\Omega(e)}\leqslant\frac{N_a(e)}{N(e)}$ for any eccentricity (respectively solid and dashed curves of Fig.\,\ref{fig:phasespace}a). As a result, the semi-major axis of most short period planets is decreasing.

On the other hand, truncating Eq.\,(\ref{evol_a}) at the order $e^2$ for the semi-major axis evolution yields
\begin{align}
\frac{1}{a}\frac{\dd a}{\dd t}=\frac{4\,a}{GM_{\star}M_p}
&\Big\{&\Kp&\left[(1+\frac{27}{2}e^2)
\frac{\op}{n}-(1+23\,e^2)\right]  \nonumber\\
&+& \Ks&\left[(1+\frac{27}{2}e^2)
\frac{\os}{n}-(1+23\,e^2)\right]\Big\}\ .
\label{evol_ae2}
\end{align}
and the previous condition becomes $\omega_\ii/n\leqslant(1+23\,e^2)/(1+\frac{27}{2}e^2)$. Up to $2^{\mathrm{nd}}$ order in eccentricity, the pseudo synchronization angular velocity is given by $\omega_\mathrm{eq}=(1+6\,e^2)n$ 
\footnote{these equations truncated at the order $e^2$ agree with equations in \S16 of \citet{FRH08}, even though they have been derived with different methods}.
One can see that $\omega_\mathrm{eq}/n=1+6\,e^2 \leqslant(1+23\,e^2)/(1+\frac{27}{2}e^2)$ only for $$e\leqslant\frac{1}{9}\sqrt{\frac{7}{2}}\approx0.208.$$ This means that even for a moderate eccentricity, $e\sim0.2$, the truncated model predicts that tides raised on a pseudo synchronous planet lead to a \textit{growth} of the semi-major axis instead of a \textit{decrease}, as obtained by the complete model. Therefore, truncating the tidal equations at the order $e^2$
for an eccentricity $e \geqslant 0.2$ not only predict quantitatively wrong but {\it qualitatively} wrong tidal evolutions. The same arguments for the evolution of the eccentricity show that tides raised on a pseudo synchronous planet lead to a \textit{growth} of the eccentricity for $$e\geqslant \sqrt{\frac{7}{66}}\approx0.326$$ and not to a \textit{decrease}. This is illustrated by Fig.\ref{fig:phasespace} that shows the pseudo synchronization curve (solid), the $\dot{a}=0$ curve (dashed) and the $\dot{e}=0$ curve (dotted) for the full model (top panel) and the truncated one (bottom panel), in the $\omega/n$ vs $e$ phase space. As demonstrated before, the pseudo synchronization curve crosses the $\dot{a}=0$ and $\dot{e}=0$ lines in the $2^{\mathrm{nd}}$ order model (Fig.\,\ref {fig:phasespace}b) whereas it does not when solving the complete Hut equations (Fig.\,\ref {fig:phasespace}a). \textit{As a result, with the truncated model, a pseudo synchronized planet can erroneously enter the zone of the phase space where its tides act to increase both the semi-major axis and the eccentricity.}
While this behavior is not observed with the constant phase lag model because it assumes that the star is slowly rotating ($\os/n\ll1$) and that the planet is near synchronization ($\op/n\approx1$) - placing them in the $\dot{a}<0$ and $\dot{e}<0$ zone of the phase space - this formal demonstration sets clear limits on the domain of validity of the quasi circular approximation.

%
%
%
%

\subsection{Underestimating tidal heating}\label{sec:edot_2ndOrder}

\begin{figure}[htbp] 
 \centering
 \resizebox{1.\hsize}{!}{\includegraphics{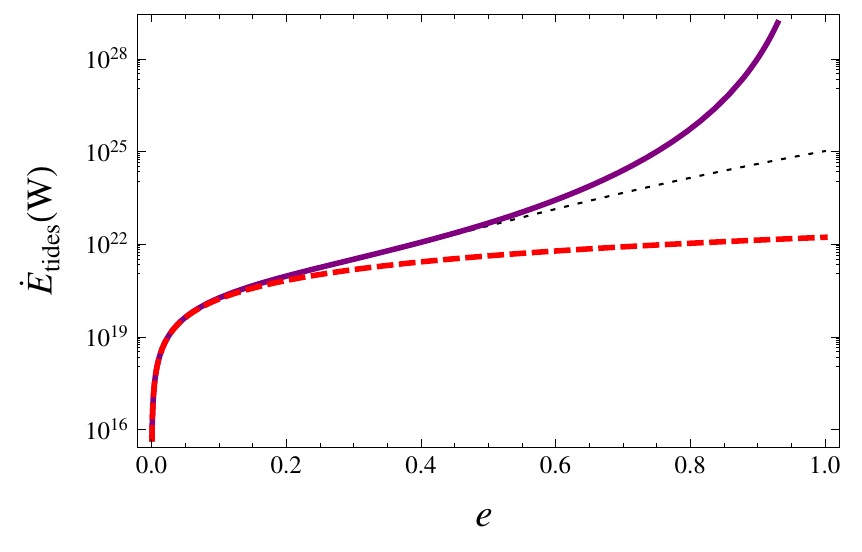}}
 \caption{Tidal energy dissipation rate in a pseudo synchronized planet (in Watt) as a function of the eccentricity calculated with Eq.\,(\ref{tidal_energy}) (\textit{solid curve}) and with the truncated formula (Eq.\,(\ref{tidal_energye2}); \textit{dashed}). The ratio of the two curves only depends on the eccentricity and not on the system's parameters. For $e=0.45$, the $e^2$ approximation (Eq.\,(\ref{tidal_energye2})) underestimates the tidal heating by a factor 10. The actual values were derived using HD 209\,458\,b parameters: $\Mp=0.657\,\mjup$, $\Rp=1.32\,\rjup$, $\Ms=1.101\,\msun$, $a=0.047\,$AU \citep{KCN07}. $Q'=10^6$ (see \S\ref{sec:q}). The \textit{dotted curve} gives the dissipation rate calculated up to $e^{10}$ (Eq.\,(\ref{tidal_energye10})).}
 \label{fig:edot}
\end{figure}

The key quantity arising from the coupling between the orbital evolution and the internal cooling history of a planet is the amount of energy dissipated by the tides in the planet's interior, which may compensate or even dominate its energy losses. As a result, tides raised in an eccentric planet can slow down its contraction \citep{BLM01,LBC09,BCB10} or even lead to a transitory phase of expansion \citep{MFJ09,ISB09}. Correctly determining
the tidal heating rate is thus a major issue in the evolution of short-period planets. The often used formula is \citep{Kau63,PC78,JGB08}
\begin{equation}
\dot{E}_{\mathrm{tides}}=
7\Kp e^2=\frac{21}{2}\frac{\kp}{Q} \left( \frac{G
\Ms^2}{\Rp} \right) \left( \frac{\Rp}{a} \right)^6 n e^2
\label{tidal_energye2}
\end{equation}
(the $21\kp/2Q$ is rewritten $63/4Q'$ in \citet{MFJ09}). As already stated by \citet{Wis08}, although this formula gives a fair approximation of the 
tidal dissipation rate for the small eccentricity cases, typical in the solar system, it severely underestimates the tidal heating for moderate
and high eccentricities. Fig.\,\ref{fig:edot} illustrates the power dissipated in a pseudo synchronized planet as a function of the eccentricity. It shows that for $e\gtrsim 0.45$, the truncated formula used in \citet{MFJ09} and \citet{ISB09} underestimates the actual tidal dissipation rate \textit{by more than one order of magnitude and by more than a factor $10^3$ for $e\gtrsim 0.7$}, an eccentricity value often advocated by these authors to explain the highly inflated planets (see \S\ref{sec:comp}).

From a mathematical point of view, the fact that a truncation to $2^{\mathrm{nd}}$ order in eccentricity yields such discrepancies is due to the presence of $(1-e^2)^{-\mathrm{15}/2}$ factors in the equations for the tidal dissipation. As already stated by \citet{Wis08}, for moderate to large eccentricity, this function is poorly represented by the first terms of its polynomial representation. Indeed, the first terms of the energy dissipation rate are given by:
\begin{equation}
\frac{\dot{E}_{\mathrm{tides}}}{7\Kp e^2}=
1+\frac{54 }{7}e^2+\frac{1133}{28} e^4+\frac{31845}{224} e^6+\frac{381909}{896} e^8+O\left(e^{10}\right).
\label{tidal_energye10}
\end{equation}
The dissipation rate calculated up to $e^{10}$ is plotted in Fig.\,\ref{fig:edot} (dotted curve), where it can be compared with the exact result. It is clear that, \textit{for $e\gtrsim0.4$, the polynomial developments of the tidal evolution equations must be done to a much higher degree that done in previous studies, or complete calculations such as the ones done in \citet{Hut81} must be used.}
The same argument holds for the evolution of the semi-major axis and the eccentricity.  Since Eqs.\,(\ref{evol_a}) and (\ref{evol_e}) also contain 
$(1-e^2)^{-\mathrm{p}/2}$ factors, the decrease of $a$ and $e$ is severely underestimated at even moderately large eccentricity when using a $2^{\mathrm{nd}}$ order
truncated expansion in eccentricity. 

In particular, as discussed in the next section, a high eccentricity ($e\gtrsim 0.6$) cannot be maintained for a few 100 Myr to a few Gyr in a system like HD 209\,458 in agreement with the results of \citet{MFJ09} (see Fig.\,\ref{fig:xo4} below).
This is in contrast with \citet{IB09} who find that the radius HD 209\,458\,b can be matched and that the system can sustain a significant eccentricity up to the observed epoch. Such discrepancies between these two studies based on the same tidal model may reveal differences in the implementations of the tidal equations, or a difference in the calculation of interior structures or boundary conditions.


\section{Effect of the truncation to $\mathbf{2^{\mathrm{nd}}}$ order in $\mathbf{e}$: Simulation results}
\label{sec:comp}

In this section, we present the comparison of the results of our complete model with the "$e^2$ model". We have calculated evolutionary tracks of the tidal evolution for various transiting systems, coupling the internal evolution of the object either with  our tidal model or with the "$e^2$ model" used in \citet{MFJ09} and \citet{IB09}. In order to ensure a consistent comparison with these authors, we directly convert their set of tidal parameters. Since our model assumes a constant time lag, and not a constant $Q'$ value, a history track computed with the $Q'$ "$e^2$ model" with a constant couple ($\qp,\,\qs$) is compared with a history track computed in our model with a constant couple $(k_2\dtp,\,k_2\dts)$ given by $(k_2\dtp=\frac{3}{2 n_{\mathrm{obs}} \qp} ,\,k_2\dts=\frac{3}{2 n_{\mathrm{obs}} \qs} )$ (See \S\ref{sec:q} and Eq.\,(\ref{QforExo})). This ensures that - although our calculations are conducted with a \textit{constant} $\Delta t$ - the quality factor computed with Eq.\,(\ref{QforExo}) in the object at the present time is the same as the one used in the \textit{$Q$ constant} model.

\subsection{Calculations at low eccentricity}

\begin{figure}[htbp]
\begin{center}
 \subfigure[Semi-major axis]{ \includegraphics[width= 4.2cm]{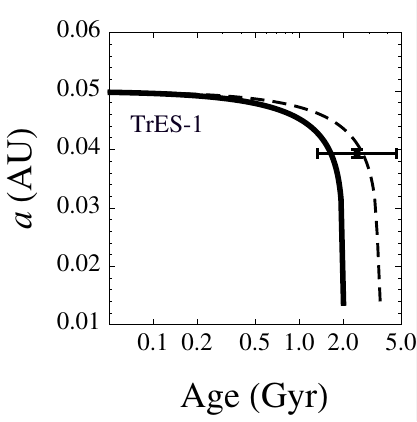} }
 \subfigure[Eccentricity]{ \includegraphics[width=4.2cm]{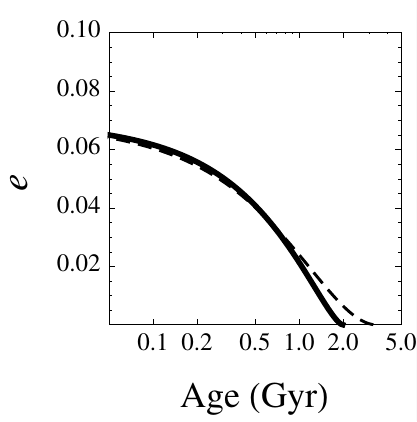}} \\
 \subfigure[Planetary radius]{ \includegraphics[width= 4.2cm]{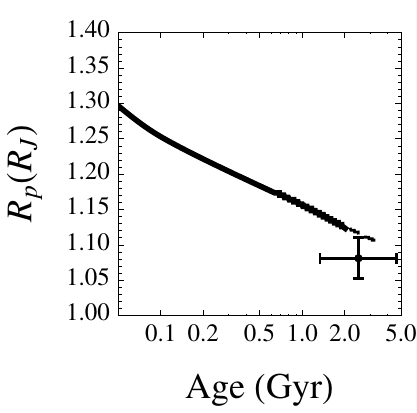} }
   \subfigure[Tidal energy dissipation]{ \includegraphics[width= 4.2cm]{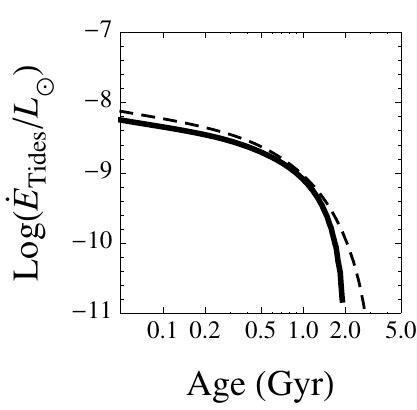} }
\end{center}
\caption{Consistent tidal/thermal evolution of TrES-1\,b computed with our constant time lag model (\textit{solid line}) and with the "$e^2$ model" (\textit{dashed line}). This is a 0.76 $\mjup$ planet orbiting
a 0.89 $\msun$ star \citep{WHR07}. The error bars are the measured parameters with the 1$\sigma$ uncertainty. The dashed curve is comparable to Fig.\,7 of \citet{MFJ09} and was computed with the same parameters ($\qp=10^{6.5}$, $\qs=10^5$). As expected, in the low $e$ limit, the tidal dissipation rate is well approximated by Eq.\,(\ref{tidal_energye2}) and the two models yield similar evolutions, although the merging time depends on the rheology used.
}
\label{fig:tres1}
\end{figure}

We first compare the results of the two models on a system which has a zero measured eccentricity and is not inflated, namely TrES-1. Such a system does not require a substantial initial eccentricity for its observed properties to be reproduced and thus provides an opportunity to test the quasi circular limit, where the "$e^2$ model" used by \citet{MFJ09} and our model should yield similar results. Fig.\,\ref{fig:tres1} illustrates the results of the integration of the coupled internal/orbital evolution equations with our constant time lag model (solid curve) and with the "$e^2$ model" (dashed curve) for an initial eccentricity of 0.07. As expected, in this low eccentricity limit both models yield very similar tracks: the eccentricity is damped to zero in a few Gyr and the semi-major axis decreases until the planet reaches the Roche limit and merges with the star, because the system does not have enough angular momentum to reach a stable equilibrium \citep{Hut80,LWC09}.
In this case, tidal heating is not sufficient to significantly affect the radius of the planet which keeps shrinking steadily as it cools. Note, however, that,
although the qualitative behavior of the evolution is the same, the hypothesis made on the rheology of the body can influence the age at which the merging occurs.

\subsection{Calculations at high eccentricity}
\label{sec:comp2}

\begin{figure}[htbp]
\begin{center}
 \subfigure[Semi-major axis]{ \includegraphics[width= 4.3cm]{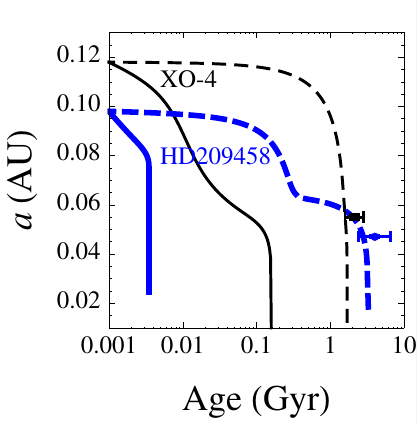} }
 \subfigure[Eccentricity]{ \includegraphics[width=4.2cm]{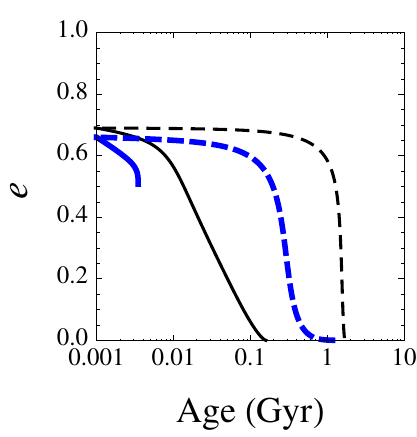}} \\
 \subfigure[Planetary radius]{ \includegraphics[width= 4.2cm]{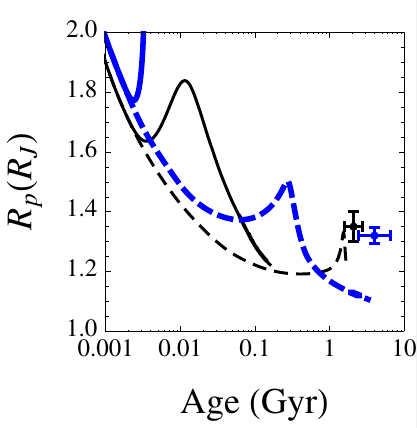} }
   \subfigure[Tidal energy dissipation]{ \includegraphics[width= 4.25cm]{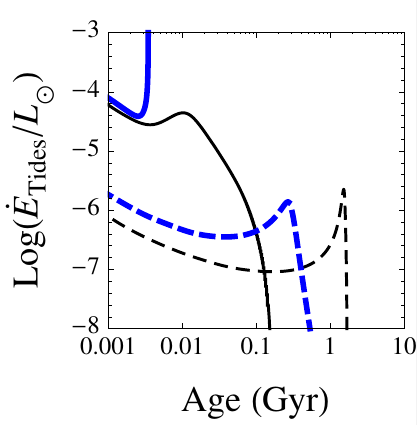} }
\end{center}
\caption{Consistent tidal/thermal evolution of XO-4\,b (\textit{thin, black}) and HD 209\,458\,b (\textit{thick, blue}) computed with our constant time lag model (\textit{solid line}) and with the "$e^2$ model" (\textit{dashed line}). XO-4\,b is a 1.72 $\mjup$ planet orbiting
a 1.32 $\msun$ star \citep{MBV08}. HD 209\,458\,b is a 0.657 $\mjup$ planet orbiting
a 1.01 $\msun$ star \citep{KCN07}. The dashed curves are comparable to Fig.\,8 and 10 of \citet{MFJ09} and were computed with the same parameters ($\qp=10^{5}$, $\qs=10^5$). In this large eccentricity regime, using the same quality factor, the "$e^2$ model" underestimates the tidal dissipation rate by 2 orders of magnitude and thus overestimates the star-planet merging timescale by a factor 10 to $10^3$.}
\label{fig:xo4}
\end{figure}

In the moderately to highly eccentric regime, the tidal dissipation rate can no longer be approximated by Eq.\,(\ref{tidal_energye2}) (see \S\ref{sec:edot_2ndOrder}). Instead, Eq.\,(\ref{tidal_energy}) must be used and yields - as shown by Fig.\,\ref{fig:edot} - a much more important dissipation rate. As a result, tidal evolution takes place on a much shorter time scale, and both the eccentricity damping and the merging with the star occur earlier in the evolution of the planet. For illustration, Fig.\,\ref{fig:xo4} portrays the possible thermal/tidal evolution (for given
initial conditions) for XO-4\,b (thin black curves) and HD 209\,458\,b (thick blue curves) computed with the "$e^2$ model" (dashed) and with our model (solid). The dashed curves are similar to the ones displayed in
 Fig.\,8 and 10 of \citet{MFJ09}. As mentioned above and illustrated in Fig.\,\ref{fig:xo4}d, the energy dissipation 
is much larger when fully accounting for the large eccentricity. The evolution of the planet can exhibit two different general behaviors:
\begin{itemize}
\item The planet first undergoes a phase of contraction and rapid cooling before the tidal heating due to the large initial eccentricity starts to dominate the energy balance of the object, leading to a phase of radius inflation (as shown by Fig.\,\ref{fig:coupling} for a test case).
\begin{figure}[htbp] 
 \centering
 \resizebox{1.\hsize}{!}{\includegraphics{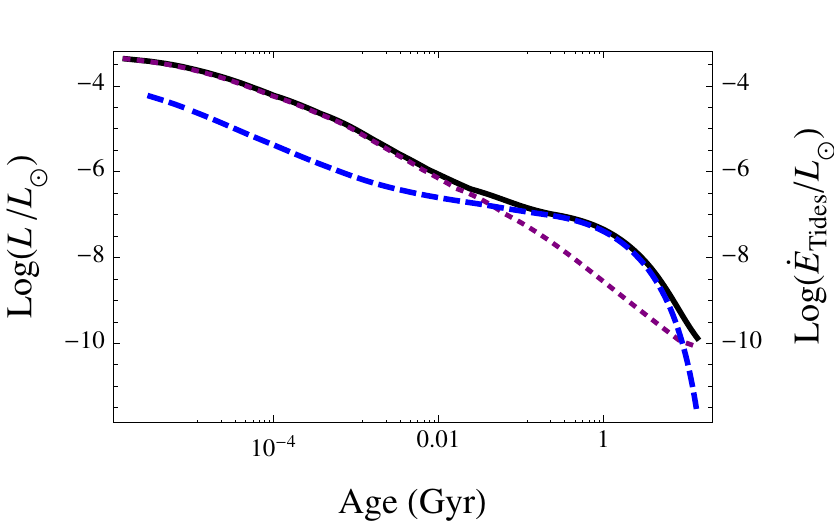}}
 \caption{Internal energy balance in the evolving planet. \textit{Solid line}: luminosity of the object with tidal heating. \textit{Dotted line}:  luminosity of the object without tidal heating. \textit{Dashed line}: tidal energy dissipation rate. The object contracts as it cools until the energy input balances its thermal losses and sustains a higher entropy in the gaseous envelop, yielding a larger radius.}
 \label{fig:coupling}
\end{figure}
 This speeds up the damping of the eccentricity and the decrease of the semi-major axis, since $\dot{a}$ and $\dot{e}\propto \Rp^5$. When the eccentricity becomes small enough, a "standard" contraction phase begins and lasts until the planet merges with the star (due to {\it stellar} tides; \citealt{LWC09}) or - if enough angular momentum is present in the system - until both tidal and thermal equilibria are achieved. Such a behavior has already been identified by \citet{MFJ09} and \citet{IB09} but, because these authors use truncated tidal equations, they find that a large eccentricity can be maintained for a few Gyr and keep inflating the planet at a late time, as illustrated on Fig.\,\ref{fig:xo4} (dashed curves); whereas it is not the case. 
 
 \item In some extreme cases, such as HD 209\,458, for the initial conditions corresponding to the ones in Fig.\,\ref{fig:xo4}, the tidal heating can overwhelm the cooling rate of the planet by orders of magnitude and lead to a spectacular inflation of the planet and thus to a rapid merging with the star. This stems from a combination of different effects. First of all, as mentioned above, the expansion of the radius accelerates the tidal evolution and thus the decrease of the orbital distance. Furthermore, the Roche limit ($a_\mathrm{R}=\alpha \Rp \sqrt[3]{\Ms / \Mp}$ where $\alpha$ is a constant which depends on the structure of the body and is equal to 2.422 for fluid objects) increases with the radius of the planet, extending the merging zone.
 \end{itemize}

As clearly illustrated by these calculations, using tidal equations truncated at second ($e^2$) order leads to severely erroneous evolutionary tracks for initially moderately ($e\gtrsim 0.2$) or highly eccentric systems. Indeed, the complete tidal model shows that, for the initial conditions and $Q$ parameter values chosen by \citet{IB09} and \citet{MFJ09}, HD 209\,458\,b would in fact have disappeared! As mentioned earlier, the use of such a quasi circular approximation cannot be
justified by the uncertainty on the quality factor, as the discrepancy in the characteristic evolution timescales can amount to 3 orders of magnitude in some cases, depending on the initial eccentricity. Conversely, trying to infer values for the stellar or planetary tidal quality factors $Q$ from tidal evolution calculations performed with the truncated $e^2$ model will lead to severely inaccurate values. 



\section{Global view of transiting systems}\label{sec:global}

As mentioned earlier, tidal heating has been suggested by several authors to explain the anomalously large radius of some giant close-in observed exoplanets. As demonstrated in \S\ref{sec:comp}, the previous calculations, all based on constant-$Q$ models truncated at the order $e^2$ yield inaccurate results when applied to significantly (initial or actual) eccentric orbits - a common situation among detected exoplanetary systems. In this section, we revisit the viability of such a tidal heating mechanism to explain the large observed Hot Jupiters radii with the present complete Hut tidal model. We first examine the properties of the known transiting systems. Then we show that, although indeed providing a possible explanation for some transiting systems, the tidal heating hypothesis fails to explain the radii of extremely bloated planets such as - among others - HD 209\,458\,b, TrES-4\,b, WASP-4\,b or WASP-12\,b, in contrast with some previously published
results based on truncated tidal models (See \S\ref{sec:disc}).

\begin{figure}[htbp] 
 \centering
 \resizebox{1.\hsize}{!}{\includegraphics{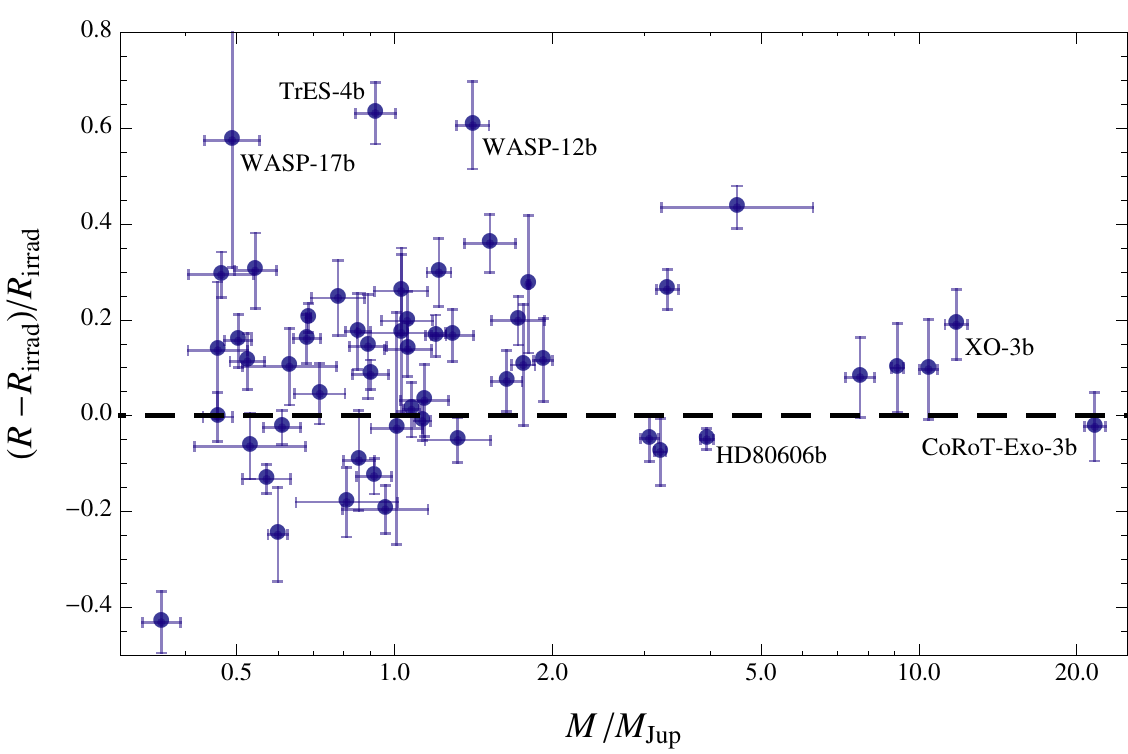}}
 \caption{Relative radius excess between the observationally and the theoretically determined values for 54 transiting systems. Objects above the dashed line are considered as being anomalously bloated compared with the prediction of the regular evolution of an irradiated gaseous planet. All the objects below this line can be explained by a heavy material enrichment in the planet's interior \citep{BCB08}.}
 \label{fig:RsurRirrad}
\end{figure}

It is now well established that a large number of transiting giant exoplanets are more inflated than predicted by the standard cooling theory of irradiated gaseous giant planets (see \citealt{US07,BCB10} for reviews).
In order to quantify this effect, we have computed the radius predicted by our standard model, described in \S\ref{sec:intevolution}, for the 54 transiting planets detected at the time of the writing of this paper, with $\Mp>0.3\mjup$ (about a Saturn mass). We define the \textit{radius excess} as the difference between the observed radius and the one predicted by the model at the estimated age of the system, denominated $R_{\mathrm{irrad}}$.
Results are summarized in Fig.\,\ref{fig:RsurRirrad}. 
The existence of objects below the $R=R_{\mathrm{irrad}}$ line is a clear signature of the presence of a dense core and/or of the enrichment of the gaseous envelop \citep{BAC06,FMB07,BHB07,BCB08,LBC09}. Note that most of the objects significantly below this line are in the $M\lesssim1\mjup$ region, and can be explained with a $M_Z/\Mp \simgr 0.10$ heavy material enrichment \citep{BCB08}, in good agreement with predictions of the core accretion scenario for planet formation (\citealt{BAC06,MAB09}).
Interestingly enough, all the planet radii in the $R\lesssim R_{\mathrm{irrad}}$ region of Fig.\,\ref{fig:RsurRirrad} show no significant eccentricity and can be explained by including the presence of a core in their internal structure and an orbital evolution with a low initial eccentricity, independently of the chosen tidal parameters.

Among the 39 remaining objects, we will focus on the most extremely inflated ones to investigate the validity of the tidal heating hypothesis to explain their abnormally low density, as they provide the most stringent cases to examine the viability of this scenario. For sake of
simplicity and to avoid introducing further free parameters in our tidal model, and since our aim is to derive an \textit{upper limit} for the radius that a planet can achieve under the effect of tidal heating, we will not consider the presence of heavy element enrichment in our calculations.

Our calculations proceed as follows: 
\begin{enumerate}
\item For each of the systems, a range of initial semi-major axis ($[a_\mathrm{i,min},a_\mathrm{i,max}]$) is found by a {\it backward integration} of the tidal equations, from present-day observed values.
\item Evolutionary tracks, coupling consistently the gravothermal evolution of the irradiated planet and the tidal heating source (Eq.\,(\ref{tidal_energy})), are then computed for $a_\ii\in[a_\mathrm{i,min},a_\mathrm{i,max}]$ and an initial eccentricity $e_\ii\in[0,0.8]$. The plausibility of such initial conditions as a remnant of early planet-disk and/or planet-planet interaction is discussed in \citet{MFJ09}.
Since total angular momentum is conserved during the tidal evolution, the {\it initial spin rate} of the star is calculated by satisfying the equality between the initial and the presently observed value of the system's total angular momentum. Calculations are performed with $Q'_{0,\star}=10^5$ and $10^6$ and $Q'_{0,\mathrm{p}}=10^6$ and $10^7$ (See \S\ref{sec:q} for a detailed discussion).
\item For each evolutionary calculation, the departure from a given measured quantity is defined as $\delta_x(t)=~\left(\frac{x(t)-x_\mathrm{p}}{\sigma_x}\right)$, where $x$ refers to $a$, $e$, $\Rp$, $\es$ or $\os$ and $\sigma_x$ to their measured uncertainty. When no error bar has been measured for the eccentricity and $e=0$ has been assumed in the light curve analysis, we take $\sigma_e=0.05$. We consider that the evolution accurately reproduces
the presently measured data if there is a time interval (compatible with the age of the system) within which \textit{all} the $\delta_x$'s are smaller than 1, meaning that each one of these parameters agrees with the measured one within 1 $\sigma$.
\end{enumerate}

\begin{figure}[htbp] 
 \centering
 \resizebox{1.\hsize}{!}{\includegraphics{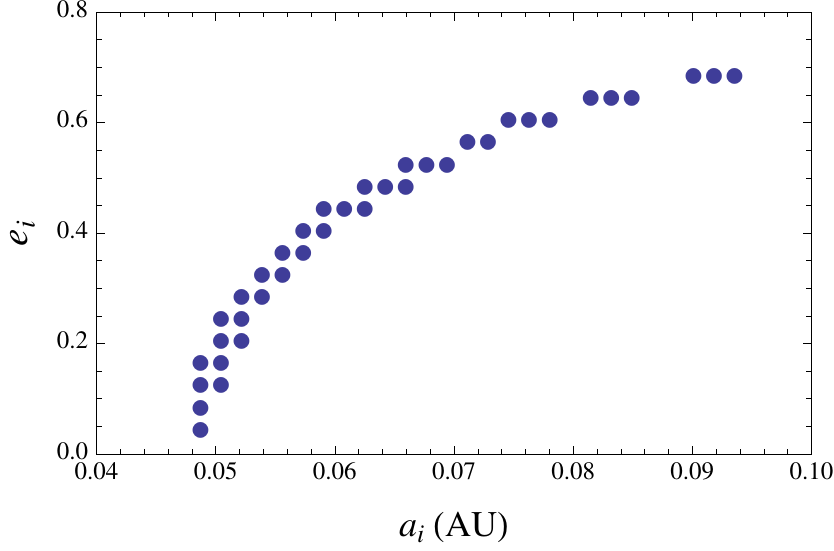}}
 \caption{Set of initial conditions yielding evolutions consistent with the actual orbital parameters of HD 209\,458\,b. These runs assume $Q'_{0,\mathrm{p}}=10^6$ and $Q'_{0,\star}=10^6$. As large eccentricity speeds up the tidal evolution, when initial eccentricity of the orbit increases, the initial semi-major axis must also increase to recover the observed parameters at the age of the system.}
 \label{fig:aiei}
\end{figure}

\begin{figure}[htbp]
\begin{center}
 \subfigure[Semi-major axis]{ \includegraphics[width= 4.35cm]{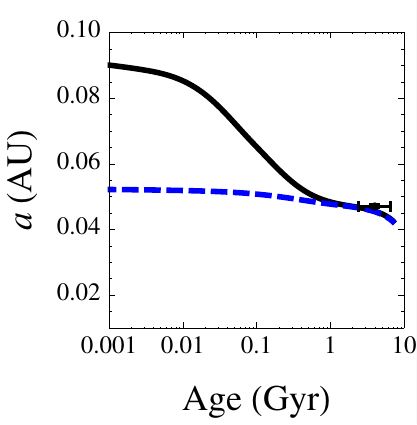} }
 \subfigure[Eccentricity]{ \includegraphics[width=4.2cm]{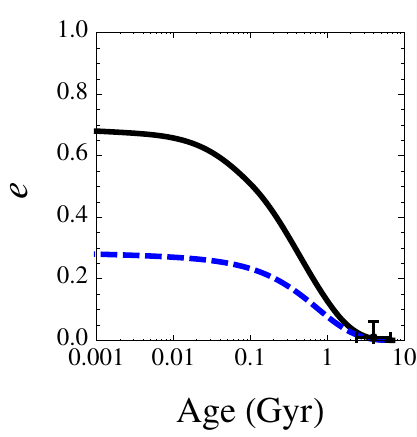}} \\
 \subfigure[Planetary radius]{ \includegraphics[width= 4.23cm]{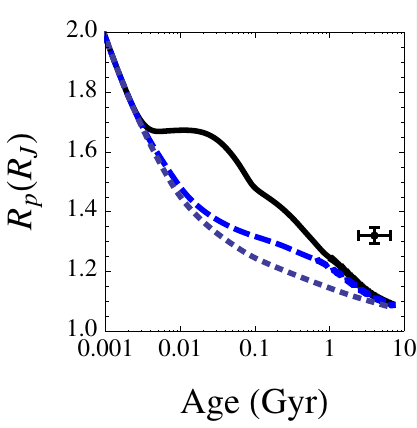} }
   \subfigure[Tidal energy dissipation]{ \includegraphics[width= 4.23cm]{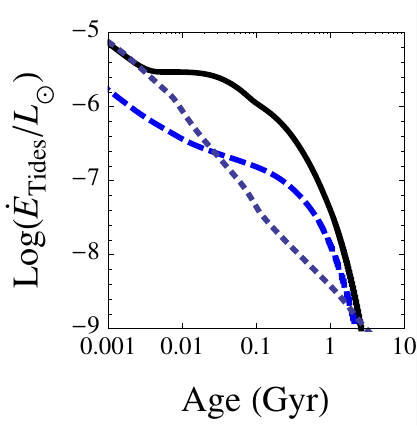} }
\end{center}
\caption{Consistent tidal/thermal evolution of HD 209\,458\,b with different initial conditions (solid and dashed) computed with our constant time lag model. HD 209\,458\,b is a 0.657 $\mjup$ planet orbiting
a 1.01 $\msun$ star \citep{KCN07}. These runs assume $Q'_{0,\mathrm{p}}=10^6$ and $Q'_{0,\star}=10^6$. For comparison, the radius and luminosity of an isolated planet (no tidal heating) is shown on the lower panels (dotted curves). Even though these evolutions recover the presently observed
orbital parameters for the system, the eccentricity damping arises too early during the evolution, leading to insufficient tidal dissipation at present epoch to explain the inflated radius.}
\label{fig:hd}
\end{figure}

\begin{figure}[htbp]
\begin{center}
  \subfigure[Semi-major axis]{ \includegraphics[width= 4.35cm]{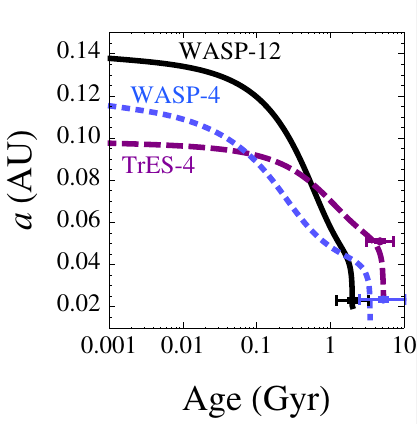} }
  \subfigure[Eccentricity]{ \includegraphics[width=4.2cm]{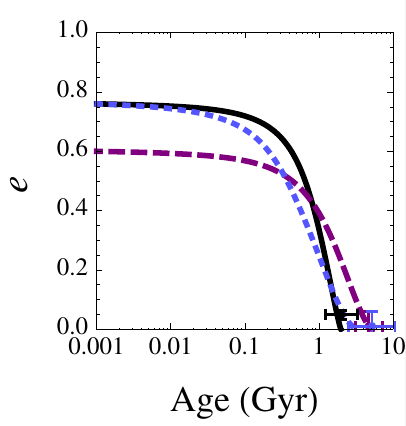}} \\
  \subfigure[Planetary radius]{ \includegraphics[width= 4.23cm]{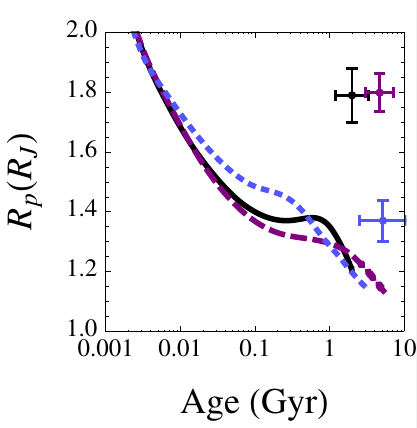} }
    \subfigure[Tidal energy dissipation]{ \includegraphics[width= 4.23cm]{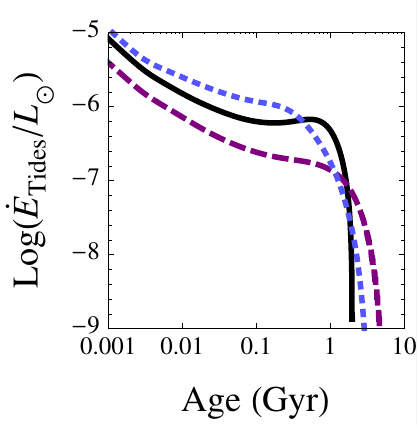} }
\end{center}
\caption{Evolutionary tracks for WASP-12\,b (solid, \citealt{HCL09}), TrES-4\,b (dashed, \citealt{DHB09}) and WASP-4\,b (dotted, \citealt{WHC09}) that lead to the best agreement with the observed orbital parameters for these systems. These runs assume $Q'_{0,\mathrm{p}}=10^6$ and $Q'_{0,\star}=10^6$. Tidal dissipation is not sufficient to sustain the large radii observed for these planets. }

\label{fig:w12_t4}
\end{figure}

Fig.\,\ref{fig:aiei} portrays a grid of evolution history initial conditions that are found to be consistent with the observed parameters of HD 209\,458, at the age of the system. As expected, an initially more eccentric system must have a greater initial separation to end up at the same location. This stems from the fact that $|\dot{a}|$ is a monotonically increasing function of $e$ for a slowly rotating star (as obtained from Eq.\,(\ref{evol_a}) for $\op=\omega_\mathrm{equ}$ and $\os/n\ll1$). Although, for these extremely bloated planets, we do find
evolutionary tracks that lead to the presently observed orbital parameters, \textit{none of these solutions can reproduce the presently observed radii}, as illustrated on Fig.\,\ref{fig:hd} for the case of HD 209\,458\,b. Indeed, the major phase of
eccentricity damping, as given by Eq.\,(\ref{evol_e}) and discussed in \S\ref{sec:comp2}, occurs too early in the evolution, so that a large fraction of the tidal heating energy dissipated in the planet has been radiated away at the age of the system ($\sim$ a few Gyrs) and can no longer provide enough energy to slow down gravitational contraction. The same behavior is found for other bloated systems such as WASP-12, TrES-4 and WASP-4 whose best evolutionary tracks are shown in Fig.\,\ref{fig:w12_t4}. For all these systems, although a solution matching the presently observed orbital parameters can be found, tidal dissipation occurs too early to reproduce the present values of the planet radii. 

\begin{figure}[htbp]
\begin{center}
  \subfigure[Semi-major axis]{ \includegraphics[width= 4.35cm]{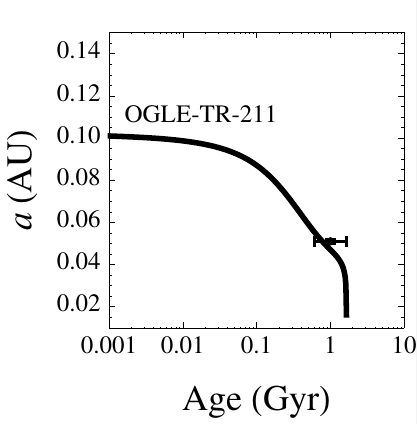} }
  \subfigure[Eccentricity]{ \includegraphics[width=4.2cm]{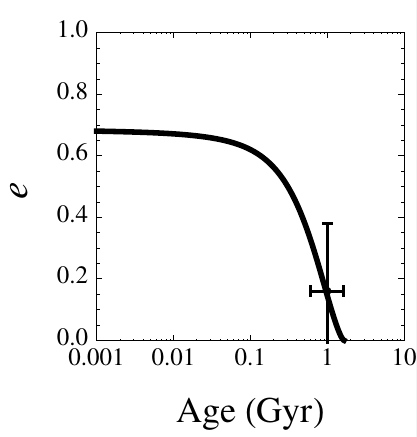}} \\
  \subfigure[Planetary radius]{ \includegraphics[width= 4.23cm]{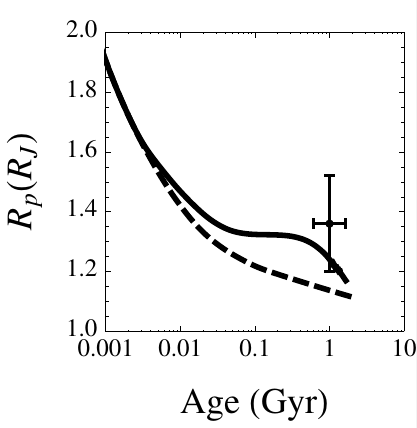} }
    \subfigure[Tidal energy dissipation]{ \includegraphics[width= 4.23cm]{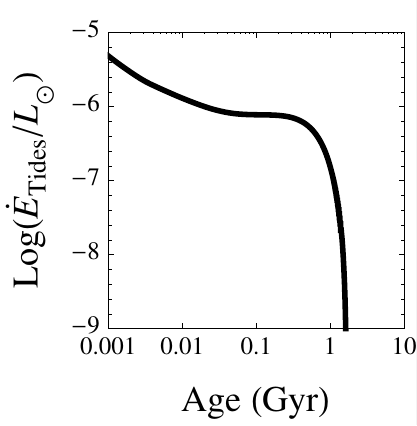} }
\end{center}
\caption{Evolutionary tracks for OGLE-TR-211\,b that lead to the best agreement with the observed parameters. is a 1.03 $\mjup$ planet orbiting
a 1.33 $\msun$ star \citep{UPN08}. These runs assume $Q'_{0,\mathrm{p}}=10^6$ and $Q'_{0,\star}=10^6$. The radius evolution of OGLE-TR-211\,b without tidal heating is plotted in panel (c) for comparison (dashed). For this moderately inflated planet, tidal heating is sufficient to sustain a large radius for the age of the system.}

\label{fig:tr211}
\end{figure}

These results, based on complete tidal evolution calculations, show that the tidal energy dissipated in the planet's tidal bulges, although providing
a viable explanation to the large radius of many short-period planets (such as OGLE-TR-211\,b shown in Fig.\,\ref{fig:tr211}), is not sufficient to explain the radii of the
most bloated planets, at the age inferred for these systems. In that case, an extra mechanism, besides tidal heating, must be invoked to solve this puzzling problem. Surface winds driven by the powerful incident stellar flux \citep{SG02}, converting kinetic energy to heat by dissipation within the tidal bulge and thus reaching deep enough layers to affect the planet's inner isentrope, or inefficient large-scale convection due
to a composition gradient \citep{CB07} could
be the other mechanisms to be considered with tidal dissipation to lead eventually to these large planet radii (see \citet{BCB10} for discussion).

\section{Discussion and conclusion}\label{sec:disc}

In this paper, we have demonstrated that the quasi-circular approximation ($e\ll 1$, i.e. tidal equations truncated at the order $e^2$) usually made in tidal calculations of transiting planet systems and valid for our Solar system planets, is not valid for the exoplanetary systems that have - or were born with - an even modestly large ($e\simgr 0.2$) eccentricity. As shown in \S\ref{sec:2ndOrder}, although the real frequency dependence of the tidal effect remains uncertain, there are dimensional evidences that for eccentric orbits, most of the tidal effect is contained in the high order terms and that truncating the tidal equations at
$2^\mathrm{nd}$ order in eccentricity can overestimate the characteristic timescales of the various orbital parameters by up to three orders of magnitude.
Therefore, truncating the tidal equations at the second
order can by no means be justified by invoking the large uncertainty in the dissipative processes and their frequency dependence. Therefore, high order tidal equations should be solved
to derive reliable results for most of the existing exoplanet transiting systems.
This need to solve the complete equations is met by any tidal model. In this context, even though no tidal model can claim describing perfectly a two body evolution, we recall that the Hut model is at least exact in the weak friction viscous approximation (see \S\ref{sec:q}).

We have tested our complete tidal model on several inflated planets to find out whether or not tidal heating can explain the large radius of most of the
 observed transiting systems. Although this mechanism is indeed found to be sufficient to explain moderately bloated planets such as OGLE-TR-211\,b (see Fig.\,\ref{fig:tr211}), we have been \textit{unable} to find evolutionary paths that reproduce both the measured radius and the orbital parameters of HD 209\,458\,b, WASP-12\,b, TrES-4\,b, and WASP-4\,b (see Fig.\,\ref{fig:hd} and Fig.\,\ref{fig:w12_t4}) for their inferred age range. The main reason is the early circularization of the orbit of these systems. As demonstrated in the paper, this stems from the non-polynomial terms in eccentricity in the complete tidal equations, which are missing when truncating the equations at small $e$-order. The present results, based on complete tidal equations, show that tidal heating,
although providing an important contribution to the planet's internal heat budget during the evolution, cannot explain {\it alone} the observed properties of all exoplanets.

This is in contrast with some
 of the conclusions reached in previous studies. Based on truncated tidal models, \citet{IB09} and \citet{ISB09} find evolutionary tracks that match observed parameters for HD 209\,458\,b, WASP-12\,b, and WASP-4\,b and thus suggest that the tidal heating is the principal cause of the large radii of Hot Jupiters.

These particular properties of Hot Jupiters, including the extreme cases of the most severely bloated planets, can only be explained if the following explanations/mechanisms occur during the system lifetimes:
\begin{itemize}
\item Early spin up of the star: simulations of the rotational evolution of solar-like stars \citep{BFA97} show that after the dispersion of the accretion disk, the rotation rate of the contracting star increases due to angular momentum conservation, until magnetic braking takes over. Considering Eq.\,(\ref{evol_e}), we see that stellar tides act as an eccentricity source if $\frac{\os}{n}\geqslant\frac{18}{11}\frac{N_e(e)}{\Omega_e(e)}$. Investigating whether the duration of this phase lasts long enough and whether the magnitude of this effect is large enough to drive enough eccentricity requires
performing consistent star/planet thermal/tidal calculations and will be investigated in a forthcoming paper.
\item Presence of a third body: as proposed by \citet{Mar07}, a low mass terrestrial planet can drive the eccentricity of a massive giant planet during
up to Gyr timescales. Accurate enough observations are necessary to support or exclude the presence of such low-mass companions.
\item As mentioned earlier, combining tidal heat dissipation with other mechanisms such as surface winds, due to the stellar insolation, dissipating deep enough in the tidal bulges, or layered convection within the planet's interior may provide the various pieces necessary to completely solve the puzzle.
\end{itemize}

In conclusion, the suggestion that tidal heating is the main mechanism responsible to solve the problem of anomalously large short-period planets, as sometimes claimed in the literature, must be reformulated more rigorously: although providing a non-negligible contribution to hot-Jupiter
heat content, tidal dissipation does not appear to provide the whole explanation. Further studies are thus necessary to eventually nail down this puzzling issue. 

\begin{acknowledgements}
This work was
supported by the Constellation european network MRTN-CT-2006-035890, the french ANR "Magnetic Protostars and Planets"
(MAPP) project and the "Programme National de Plan\'etologie" (PNP) of CNRS/INSU. We acknowledge the use of the \textit{www.exoplanet.eu} database. We thank our referee, J. Fortney, for helpful suggestions.
\end{acknowledgements}


\bibliography{biblio} 

\bibliographystyle{aa} 

\clearpage
\appendix
\section{Tidal evolution equations for finite eccentricity and obliquity.}
\label{appendix}

The present calculation of the tidal evolution equations extends the formulas given in \citet{Hut81} to any obliquity. We consider a system of two deformable bodies of mass $M_1$ and $M_2$.
The demonstration follows three main steps. First we compute a vector expression for the tidal force and torque. Second we derive the variation of the rotation rate, obliquity and orbital angular momentum thanks to this expression of the torque and using the \textit{total} angular momentum conservation. Finally the evolution of the semi-major axis and eccentricity are obtained from the expression of the energy dissipated by tides in the deformable body. The total amount of energy dissipated by tides in one of the bodies is a direct product of the calculation.

 Up to the quadrupolar terms in the tidal deformation, the mutual interaction of the tidal bulges is negligible and we can consider separately the effects of the tides raised in each body and sum them up at the end of the calculation. Let us consider the effect of the tides raised in a deformable body (say $M_1$, hereafter the primary) in interaction with a point mass (say $M_2$ the secondary). The mass distribution of a deformable body in a quadrupolar tidal potential can be mimicked by a central mass $M_1-2m$ and two point masses at the location of the tidal bulges ($\mathbf{r_+,\,r_-}$) of mass $m$ with $\|\mathbf{r_+}\|=\|\mathbf{r_-}\|=R_1$ the radius of the primary and $m=\frac{1}{2}k_2 M_2\left(\frac{R_1}{r}\right)^3$ where $k_2$ is the love number of degree 2 of the primary and $r$ is the distance between the center of the two objects. Since we consider a \textit{constant} time lag $\Delta t_1$ between the deforming potential and the tidal deformation in the frame rotating with the primary, $\mathbf{\hat{r}_+}=\mathbf{\hat{r}}(t-\Delta t_1)$ ( $\mathbf{\hat{r}}$ refers to the unit vector associated to $\mathbf{r}$) in this frame. Let $\mathbf{\dot{\theta}}$ be the orbital rotation vector collinear to the orbital angular momentum and whose value is the instantaneous variation rate of the true anomaly $\theta$ of the bodies in their keplerian motion and  $\mathbf{\omega_1}$ the rotation vector of the primary. Thus, to first order in $\Delta t_1$, 
\begin{equation}
\mathbf{\hat{r}_+}=\mathbf{\hat{r}}(t-\Delta t_1)\approx \mathbf{\hat{r}}-\Delta t_1\,\mathbf{\hat{r}}\times (\mathbf{\omega_1}-\mathbf{\dot{\theta}}),
\end{equation}
the amplitude of the tidal bulges also lags behind the deforming potential and is given by 
\begin{align}
m(t)\ \ \ &=&\frac{1}{2}k_2 M_2&\left(\frac{R_1}{r(t-\Delta t_1)}\right)^3 \nonumber \\
&\approx &\frac{1}{2}k_2 M_2&\left(\frac{R_1}{r}\right)^3(1+3\frac{\dot{r}}{r}\Delta t_1) ,
\end{align}
and the force exerted by this  mass distribution on the secondary is
 \begin{align}
 \label{tideForce}
    \mathbf{F}=-\ \frac{GM_1 M_2}{r^2}\cdot \mathbf{\hat{r}} \ \ &-&3\frac{Gk_2M_2^2R_1^5}{r^7}&\{1+3\frac{\dot{r}}{r}\Delta t_1\}\cdot\mathbf{\hat{r}}\nonumber\\
   &+&3\frac{Gk_2M_2^2R_1^5}{r^7}& \Delta t_1(\mathbf{\omega_1}-\mathbf{\dot{\theta}}) \times \mathbf{\hat{r}}.
\end{align}
Thus the tidal torque reads:
  \begin{equation}
  \label{torque}
    \mathbf{N}=3\frac{Gk_2M_2^2R_1^5}{r^6}\, \mathbf{\hat{r}}\times\left(\Delta t_1(\mathbf{\omega_1}-\mathbf{\dot{\theta}}) \times \mathbf{\hat{r}}\right).
 \end{equation}
and the angular momentum conservation yields
   \begin{equation}
   \label{angcons}
    \mathbf{N}=\mathbf{\dot{h}}=-\mathbf{\dot{L}},
 \end{equation}
where $\mathbf{h}=\frac{M_1M_2}{M_1+M_2}na^2\sqrt{1-e^2}$ is the orbital angular momentum and $\mathbf{L}=C_1\omega_1$ the rotational angular momentum of the primary. We can then simply derive the rate of angular velocity variation:
   \begin{equation}
   \label{comedot}
\frac{\dd}{\dd t}(C_1\omega_1)=\dot{L}=\mathbf{\dot{L}}\cdot\mathbf{\hat{L}}=-\mathbf{N}\cdot\mathbf{\hat{L}}
 \end{equation}
 This product can be carried out by projecting in any base. We choose the base defined by $\mathbf{h}=(0,\,0,\,h)$ and $\mathbf{\omega_1}=(\omega_1 \sin \varepsilon_1,\,0,\,\omega_1 \cos \varepsilon_1)$ where $\varepsilon_1$ is the obliquity. In this base, 
    \begin{align}
    \mathbf{N}=3\frac{Gk_2M_2^2R_1^5}{r^6}\Delta t_1 \left( \begin{array}{c} \omega_1 \sin \varepsilon_1 \sin^2(\theta+\psi) \\ -\omega_1\sin \varepsilon_1 \cos(\theta+\psi)\sin(\theta+\psi) \\ \omega_1\cos \varepsilon_1-\dot{\theta} \end{array}\right),
 \end{align}
 where $\psi$ is the longitude of the periapsis in this base. The precession of the periapsis occurring on a much shorter timescale than the tidal migration, we can average the tidal torque over $\psi$. This yields:
     \begin{align}
          \label{averagetorque}
    \mathbf{N}=3\frac{Gk_2M_2^2R_1^5}{r^6}\Delta t_1 \left( \begin{array}{c} \frac{1}{2}\omega_1 \sin \varepsilon_1  \\ 0 \\ \omega_1\cos \varepsilon_1-\dot{\theta} \end{array}\right).
 \end{align}
 We can compute the dot product in Eq.\,(\ref{comedot}) giving (with $x_1=\cos \varepsilon_1$)
\begin{equation}
\label{dcomeg}
\frac{\dd C_1\omega_1}{\dd t}=3\frac{Gk_2 \Delta t_1 M_2^2R_1^5}{ r^6} \left(x_1 \dot{\theta}-\left(\frac{1+x_1^2}{2} \right) \omega_1 \right).
\end{equation} 
The mean rotation rate variation (Eq.\,(\ref{rot_tidal})) is obtained by averaging over a keplerian orbit using
\begin{equation}
r=a\frac{1-e^2}{1+e\cos\theta},
  \end{equation}
\begin{equation}
\label{average_r6}
 \frac{1}{T_\mathrm{orb}}  \oint_{\mathrm{orbit}}\left(\frac{a}{ r}\right)^6 \cdot \dd t =\oint_{\mathrm{orbit}}\frac{a^6}{\dot{\theta} r^6} \cdot \dd \theta =\Omega(e)
  \end{equation} 
  and
 \begin{equation}
\label{average_thetar6}
 \frac{1}{T_\mathrm{orb}}  \oint_{\mathrm{orbit}}\dot{\theta}\left(\frac{a}{ r}\right)^6 \cdot \dd t =n\,N(e)
  \end{equation}  
 where $\dd t=\dd \theta/\dot{\theta}$, and the angular momentum conservation over one orbit is used to express $\dot{\theta}$ with respect to $\theta$ (see Eq.\,(\ref{omega_e}) and Eq.\,(\ref{n_e}) for the definition of $\Omega(e)$ and $N(e)$).
The variation of the obliquity can be obtained with:
\begin{equation}
\dot{x}_1=\dot{(\mathbf{\hat{L}}\cdot\mathbf{\hat{h}})}\,=\mathbf{\dot{\hat{L}}}\cdot\mathbf{\hat{h}}+\mathbf{\hat{L}}\cdot\mathbf{\dot{\hat{h}}}.
\end{equation} 
Carrying out the differentiation and using Eq.\,(\ref{angcons}) yields
\begin{align}
\dot{x}_1&=&\frac{(\mathbf{\hat{L}}\cdot\mathbf{\hat{h}})(\mathbf{N}\cdot\mathbf{L})}{L^2}-\frac{(\mathbf{N}\cdot\mathbf{\hat{h}})}{L}\nonumber\\
&&-\frac{(\mathbf{\hat{h}}\cdot\mathbf{\hat{L}})(\mathbf{N}\cdot\mathbf{h})}{h^2}+\frac{(\mathbf{N}\cdot\mathbf{\hat{L}})}{h}.
\end{align} 
Subsituting Eq.\,(\ref{averagetorque}) for $\mathbf{N}$ we get after simplification
\begin{equation}
\frac{\dd \varepsilon_1}{\dd t}=\frac{3}{2}\frac{Gk_2 \Delta t_1 M_2^2R_1^5}{ r^6} \sin \varepsilon_1\left(\frac{x_1}{C_1}-2\frac{\dot{\theta}}{C_1\omega_1}-\frac{\omega_1}{h}  \right).
\end{equation} 
Averaging over an orbit using Eqs.\,(\ref{average_r6})-(\ref{average_thetar6}) gives Eq.\,(\ref{rot_tidal2}).

To obtain the variation of the semi-major axis and eccentricity, we must compute the work done by the tidal force on the secondary:
  \begin{equation}
  < \dot{E}_\mathrm{orb}>=\frac{1}{T_\mathrm{orb}}  \oint_{\mathrm{orbit}}\mathbf{F}\cdot \dd \mathbf{r},
  \end{equation}
  \begin{equation}
  < \dot{E}_\mathrm{orb}>=\frac{1}{T_\mathrm{orb}} \oint_{\mathrm{orbit}} (\,\dot{r}\,\mathrm{F}_r\,+\mathbf{N\cdot\dot{\theta}})\,\dd t ,
\end{equation}
where $\mathrm{F}_{r} $ is the radial component and $\mathbf{N\cdot\dot{\theta}}$ the normal one. The radial forces in $r^{-2}$ and $r^{-7}$ in Eq.\,(\ref{tideForce}) are conservative and do not dissipate energy. The radial part of the work is computed using
     \begin{equation}
\dot{r}=an\frac{e}{\sqrt{1-e^2}} \sin \theta,
  \end{equation}
  and thus
\begin{equation}
\label{average_r6rpoint}
 \frac{1}{T_\mathrm{orb}}  \oint_{\mathrm{orbit}}\left(\frac{a}{ r}\right)^8\left(\frac{\dot r}{ a}\right)^2 \cdot \dd t =\frac{n^2e^2}{2}\frac{N_e(e)}{1-e^2}
   \end{equation} 
(see Eq.\,(\ref{ne_e}) for the definition of $N_e(e)$). The normal component can be written
\begin{equation}
\mathbf{N\cdot\dot{\theta}}\,\dd t=3Gk_2 \Delta t_1 M_2^2R_1^5 \left( \frac{x_1 \omega_1 -\dot{\theta}}{r^6}\right)\dd \theta.
\end{equation}
Again, averaging is carried out using Eqs.\,(\ref{average_r6})-(\ref{average_thetar6}). After integration,
\begin{equation}
\label{appendtidalorbnrj}
  < \dot{E}_\mathrm{orb}>=2K_1\left[N(e)\,x_\mathrm{1}\,\frac{\omega_1}{n} -N_a(e)\right].
\end{equation}
The variation of semi-major axis due to the tides raised in the primary (Eq.\,(\ref{evol_a})) is directly given by
\begin{equation}
  < \dot{E}_\mathrm{orb}>=-\frac{\dd}{\dd t}\frac{GM_1M_2}{2\,a}=\frac{GM_1M_2}{2\,a^2}\dot{a}
  \end{equation}
Since the orbital angular momentum is given by
   \begin{equation}
   \label{orbangmom}
   h=\sqrt{G\frac{M_1^2M_2^2}{M_1+M_2}a(1-e^2)}.
  \end{equation}
the variation of the eccentricity can be obtained by differentiating $h$ with respect to $t$:
\begin{equation}
   \label{varorbangmom}
\frac{2\dot{h}}{h}=\frac{\dot{a}}{a}-\frac{2e\dot{e}}{1-e^2}.
\end{equation} 
Only \textit{total} angular momentum is conserved, then $\dot{h}=-\dd(C_1\omega_1)/\dd t$ and substituting Eq.\,(\ref{evol_a}),  (\ref{rot_tidal}) and (\ref{orbangmom}) in (\ref{varorbangmom}) gives one of the two terms of Eq.\,(\ref{evol_e}) corresponding to the contribution of one
of the two bodies (star or planet) for the evolution of the eccentricity.
Finally, the rate of tidal energy dissipation into the primary is
\begin{equation}
\dot{E}_{\mathrm{tides}}=\frac{\dd}{\dd t}\frac{GM_1M_2}{2\,a}-\omega_1\frac{\dd}{\dd t}(C_1\omega_1).
\label{appendtidalnrj}
\end{equation}
Thus, substituting Eqs.\, (\ref{dcomeg}) and (\ref{appendtidalorbnrj}) in Eq.\, (\ref{appendtidalnrj}) gives
\begin{equation}
\dot{E}_{\mathrm{tides}}=2K_1\left[N_a(e)-2N(e)\,x_\mathrm{1}\,\frac{\omega_1}{n} +\left(\frac{1+x_1^2}{2}\right)\Omega(e)\left(\frac{\omega_1}{n}\right)^2\right] .
\label{appendtidalnrj2}
\end{equation}
One can see that the dissipated energy is positive for any value of $e$ and $x_1$ as expected \citep{Hut81}, and that it is minimum when the body is pseudo synchronized. Substituting $\omega_1$ by the pseudo synchronization rate (Eq.\,(\ref{rot_eq})), Eq.\, (\ref{appendtidalnrj2}) simplifies to Eq.\, (\ref{tidal_energy}) that can be used for a close-in gas-giant exoplanet. For rocky planets locked in synchronous rotation by their permanent quadrupolar mass distribution, the heating rate can be estimated setting $\omega_1=n$ in Eq.\, (\ref{appendtidalnrj2}).

\textit{In fine}, the complete equations taking into account tides in both bodies are obtained by computing the effects of the tides raised in the secondary (given by the same equations with $1\rightleftarrows2$) and by adding them up to the effects of the tides in the primary.


\end{document}